\documentclass[aip,amsmath,reprint,amssymb]{revtex4-2}
\usepackage[utf8]{inputenc}
\usepackage[T1]{fontenc}
\usepackage{lmodern}
\usepackage{graphicx}
\usepackage{bm}
\usepackage{mathptmx}
\usepackage{mathrsfs}
\usepackage[hidelinks]{hyperref}
\hypersetup{
  pdftitle={Hierarchical Control of Antiphase Synchronization in Biomorphic Spiking Neural Architectures for Locomotor Control},
  pdfauthor={Artem Guba; Nikolai Kovalev; Alexey Zharinov; Vyacheslav Demin; Viktor B. Kazantsev}
}

\begin{document}

\title{Hierarchical Control of Antiphase Synchronization in Biomorphic Spiking Neural Architectures for Locomotor Control}

\author{Artem Guba}
\email{guba.av@mipt.ru}
\affiliation{Moscow Institute of Physics and Technology, 9 Institutskiy per, Dolgoprudny, 141701, Moscow Region, Russia}
\affiliation{Lobachevsky State University of Nizhny Novgorod, 23 Gagarin Avenue, Nizhny Novgorod, 603950, Russia}
\affiliation{National Research Centre Kurchatov Institute, 1 Kurchatov Square, Moscow, 123182, Russia}

\author{Nikolai Kovalev}
\email{kovalev.na@pystech.edu}
\affiliation{Moscow Institute of Physics and Technology, 9 Institutskiy per, Dolgoprudny, 141701, Moscow Region, Russia}
\affiliation{Lobachevsky State University of Nizhny Novgorod, 23 Gagarin Avenue, Nizhny Novgorod, 603950, Russia}

\author{Alexey Zharinov}
\email{zharinov@neuro.nnov.ru}
\affiliation{Moscow Institute of Physics and Technology, 9 Institutskiy per, Dolgoprudny, 141701, Moscow Region, Russia}
\affiliation{Lobachevsky State University of Nizhny Novgorod, 23 Gagarin Avenue, Nizhny Novgorod, 603950, Russia}

\author{Vyacheslav Demin}
\email{demin.vyacheslav@mail.ru}
\affiliation{National Research Centre Kurchatov Institute, 1 Kurchatov Square, Moscow, 123182, Russia}

\author{Viktor B. Kazantsev}
\email{kazantsev.vb@mipt.ru}
\affiliation{Moscow Institute of Physics and Technology, 9 Institutskiy per, Dolgoprudny, 141701, Moscow Region, Russia}
\affiliation{Lobachevsky State University of Nizhny Novgorod, 23 Gagarin Avenue, Nizhny Novgorod, 603950, Russia}
\affiliation{Skolkovo Institute of Science and Technology, Bolshoy Boulevard, 30, bld.1, Moscow, 121205, Russia}

\date{17 November 2026}

\begin{abstract}
Central pattern generators (CPGs) provide a biologically inspired foundation for adaptive locomotor control, yet systematically characterizing their dynamic regimes in spiking neural networks remains challenging. This study introduces a framework for identifying and stabilizing antiphase synchronization in a biomorphic spiking CPG model embedded in a closed neuromuscular loop. A composite synchronization metric is developed, integrating phase coherence, amplitude balance, and temporal stability. Its performance is evaluated across systematically varied stimulation waveforms and synaptic configurations. Statistical analysis reveals that stimulation type dominates regime determination, with tonic drive yielding the most robust antiphase synchronization. Periodic inputs induce nonlinear saturation and context-dependent network routing. Although synaptic weights alone show limited discriminative power, they significantly modulate robustness within a given stimulation context, with reciprocal Ia interneuron connections emerging as key stabilizers. The proposed composite metric achieves robust antiphase detection, outperforming traditional phase-only measures while preserving biophysical interpretability. These findings establish a statistically validated, hierarchical modeling approach in which external drive defines the operating regime and targeted synaptic configurations fine-tune dynamic stability, providing actionable guidelines for closed-loop neuromechanical controllers and adaptive neuroprosthetic interfaces.
\end{abstract}


\maketitle

\section{Introduction}
\label{sec1}

Central pattern generators (CPGs) are specialized neural circuits capable of producing the rhythmic activity underlying locomotion, respiration, and mastication in vertebrates \cite{Marder2001}. Because CPGs intrinsically generate stable, phase-locked oscillations, they have become highly attractive templates for bio-inspired robotic control \cite{Ijspeert2008}, offering adaptive coordination, energy efficiency, and seamless integration of sensory feedback \cite{Thandiackal2021}.

CPG models span a wide spectrum of biological detail, both in the richness of their constituent elements and in their network topology --- factors that fundamentally shape their capabilities and limitations. The most common approach treats CPGs as systems of coupled nonlinear oscillators. At the simplest end of this spectrum lie the classical Matsuoka models \cite{Matsuoka1987}, in which mutual inhibition between pairs of neurons gives rise to stable antiphase oscillations \cite{Matsuoka1985}. More elaborate architectures incorporate generalized phase generators \cite{Righetti2006a} or closed-loop hierarchical networks capable of supporting complex phase relationships \cite{Szczecinski2017}. These oscillator-based models can faithfully reproduce key locomotor patterns-from gait transitions in salamanders and swimming-to-crawling switches in fish to multimodal locomotion in amphibious robots \cite{Ijspeert2007,Crespi2008,Yu2013}. Crucially, however, their parameters are typically tuned empirically to match a desired target pattern, offering little insight into the neuronal mechanisms responsible for rhythm generation \cite{Sproewitz2008}.

Existing CPG models span a wide spectrum, from systems of coupled nonlinear oscillators \cite{Matsuoka1987} to biophysically detailed spiking neural networks \cite{Zharinov2022}. Oscillator-based models successfully reproduce key locomotor patterns \cite{Ijspeert2007,Yu2013}, however, their parameters are tuned empirically to match a target pattern, offering little insight into the neuronal mechanisms of rhythmogenesis \cite{Sproewitz2008}. Spiking models provide a powerful alternative, granting direct access to how cellular properties, synaptic plasticity, and network architecture collectively shape emergent dynamics. Among these, the Izhikevich formalism \cite{Izhikevich2003} stands out, capturing a wide repertoire of dynamic regimes exhibited by real neurons while remaining computationally efficient enough for large-scale network simulations \cite{Izhikevich2008}. Modular architectures deserve particular attention, as they introduce additional behavioral flexibility and richer emergent dynamics \cite{Rybak2015,Rybak2006}. Another significant advantage of spiking CPGs is the possibility of realizing them on specialized neuromorphic platforms, which are highly effective, compact, possess high performance and suitable for edge computing \cite{Mikhaylov2020,Gordleeva2024,Alyaev2025}.

Despite their advantages, spiking CPGs present a serious methodological challenge. The majority of studies focus on finding a single parameter set that produces a single target regime \cite{Wang2019,Korkmaz2021}.Attempts to explore multiple regimes have relied either on bifurcation analysis, which requires exact knowledge of the model \cite{Serrano2024}, or on specialized methods tied to particular neuromorphic platforms \cite{Angelidis2021,RostroGonzalez2015}. The situation becomes considerably more challenging when moving beyond single parameter sets, as the high dimensionality of the parameter space makes exhaustive exploration impractical. Additional difficulties arise from the inherent nonlinearity, multistability, and fundamental complexity of interpreting the dynamics of such networks \cite{Ruiz2024,Calim2018}.A further open question concerns how dynamic regimes are distributed in systems that couple a spiking CPG with a muscle actuator. Although the two-level Rybak architecture \cite{Rybak2015} provides a biologically plausible substrate for close-loop CPG onto a muscle output, no systematic mapping of muscle activation regimes in such systems has yet been carried out.

In experimental neuroscience, Phase Locking Value (PLV) is widely used to quantify synchronization within neural ensembles \cite{Lachaux1999}. PLV is a robust metric to amplitude fluctuations and capable of detecting synchronization in noisy, non-stationary signals \cite{Celka2007}. PLV quantifies the consistency of the phase relationship between two signals, ranging from 0 (no phase synchronization) to 1 (perfect phase locking). In addition, it can detect intermittent synchronization, which is particularly important when analyzing the non-stationary spiking dynamics typical of biological and bio-inspired neural networks. The method has been successfully deployed in applications ranging from analysis of functional connectivity in EEG recordings \cite{Selvamani2026} to the study of synchronization in model spiking networks \cite{Lowet2016}. Despite this versatility, it has not yet been adopted for the classification of operating regimes in spiking CPGs.

The present work provides the first systematic description of the dynamic regimes that arise in a closed-loop neuromuscular system built on the Rybak architecture \cite{Rybak2015}. The goal is to establish which types of muscle activation --- from the complete absence of rhythmic activity to stable antiphase synchronization --- such a system can produce when synaptic coupling strengths and the parameters of external stimulation (amplitude, frequency, and waveform) are varied. The study was carried out using our SpikingNN simulation environment and a quantitative metric based on PLV.

The following results are reported in our study: a specialized simulation environment SpikingNN for closed-loop neuromuscles CPG modeling was developed; a method for quantitative mapping of dynamic regimes based on the PLV matrix was proposed and validated; a large-scale annotated dataset comprising over $10^4$ simulation runs with systematically varied synaptic and stimulation parameters was generated and made publicly available; key parametric drivers governing regime transitions and the stability of antiphase synchronization were identified using interpretable machine learning techniques; and the proposed approach was benchmarked against classical gradient-boosting algorithms.

\section{Materials and Methods}
\label{sec:methods}

\begin{figure*}[t]
    \centering
    \includegraphics[width=0.92\textwidth]{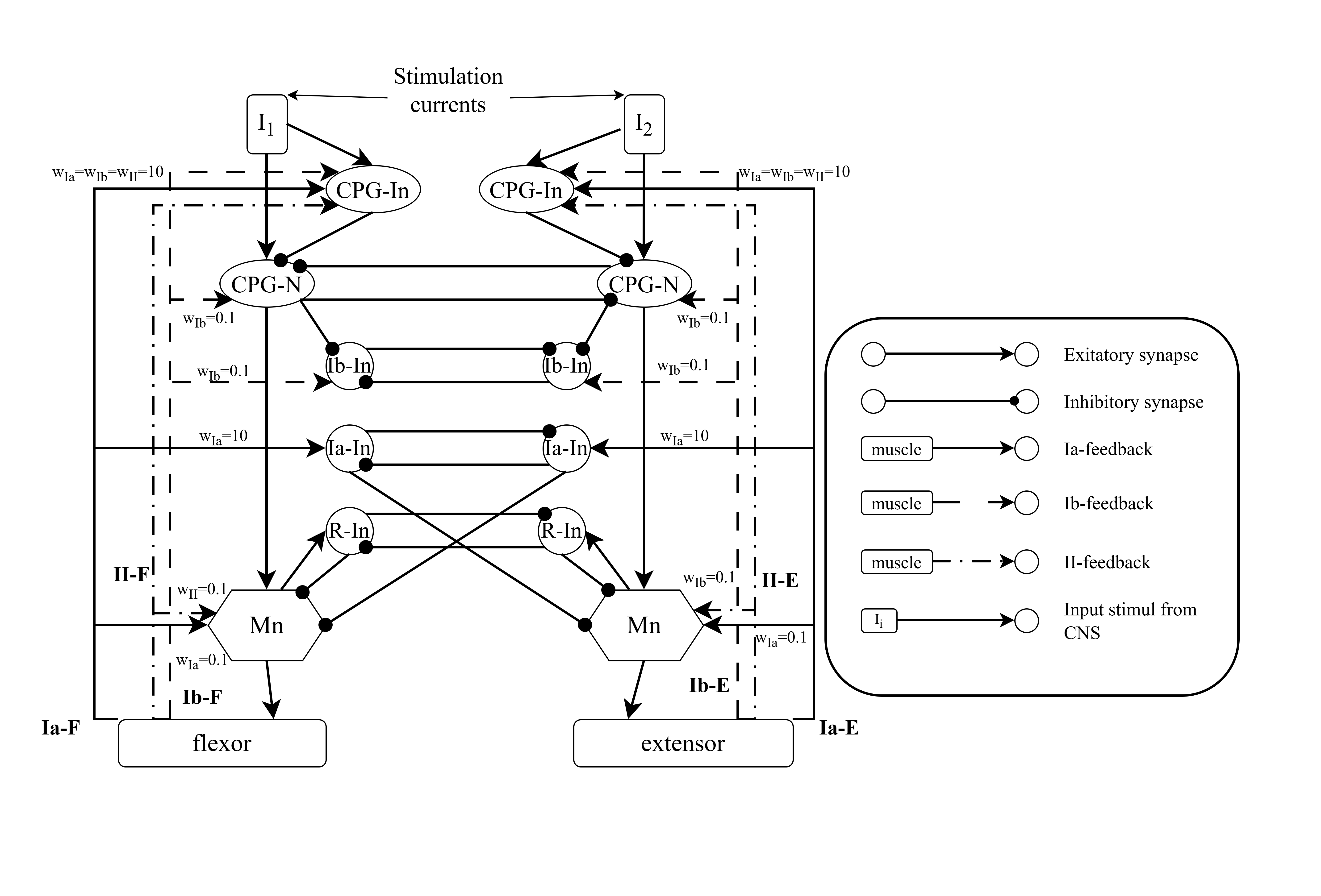}
    \caption{Schematic diagram of the CPG architecture adapted from Rybak et al. \cite{Rybak2002}. The network consists of reciprocal inhibition circuits generating rhythmic bursting (CPG-N), afferent feedback pathways (Ia, Ib), and interneurons modulating afferent input (Ib-In, Ia-In, R-In).}
    \label{fig:cpg}
\end{figure*}

The architecture of the central pattern generator (CPG) employed in this study is grounded in the half-center oscillator framework proposed by Rybak et al. \cite{Rybak2002}. As illustrated in Fig.~\ref{fig:cpg}, the model comprises two reciprocally coupled Neuronal Modules (NMs), each dedicated to the control of a single antagonistic muscle group (flexor and extensor). These modules integrate intrinsic rhythmogenesis with spinal reflex pathways to produce adaptive locomotor output.

Each neural module contains specialized neurons with distinct functions. The $\alpha$-motoneuron(Mn) serves as the final output, directly activating muscles. The Renshaw(R-In) interneuron receives feedback from the motoneuron and inhibits it, preventing excessive firing and helping to terminate motor bursts. Proprioceptive control is handled by Ia and Ib(Ia-In, Ib-In) interneurons: Ia interneurons integrate signals about muscle length and velocity, while Ib interneurons process muscle force. Together, these circuits continuously adjust motor commands in response to mechanical feedback.

Rhythmogenesis and phase switching are orchestrated by two core CPG components: the principal bursting neuron (CPG-N) and the fast-spiking switching interneuron (CPG-In). The CPG-N provides tonic excitatory drive to the $\alpha$-MN, sustaining the locomotor rhythm. The CPG-In regulates phase transitions by reciprocally inhibiting the contralateral CPG-N, ensuring alternating activation of flexor and extensor modules. This architecture operates in concert with descending commands and peripheral feedback, yielding robust, self-sustaining, and environmentally adaptive locomotor patterns.

\begin{figure*}[t]
    \centering
    \includegraphics[width=\textwidth]{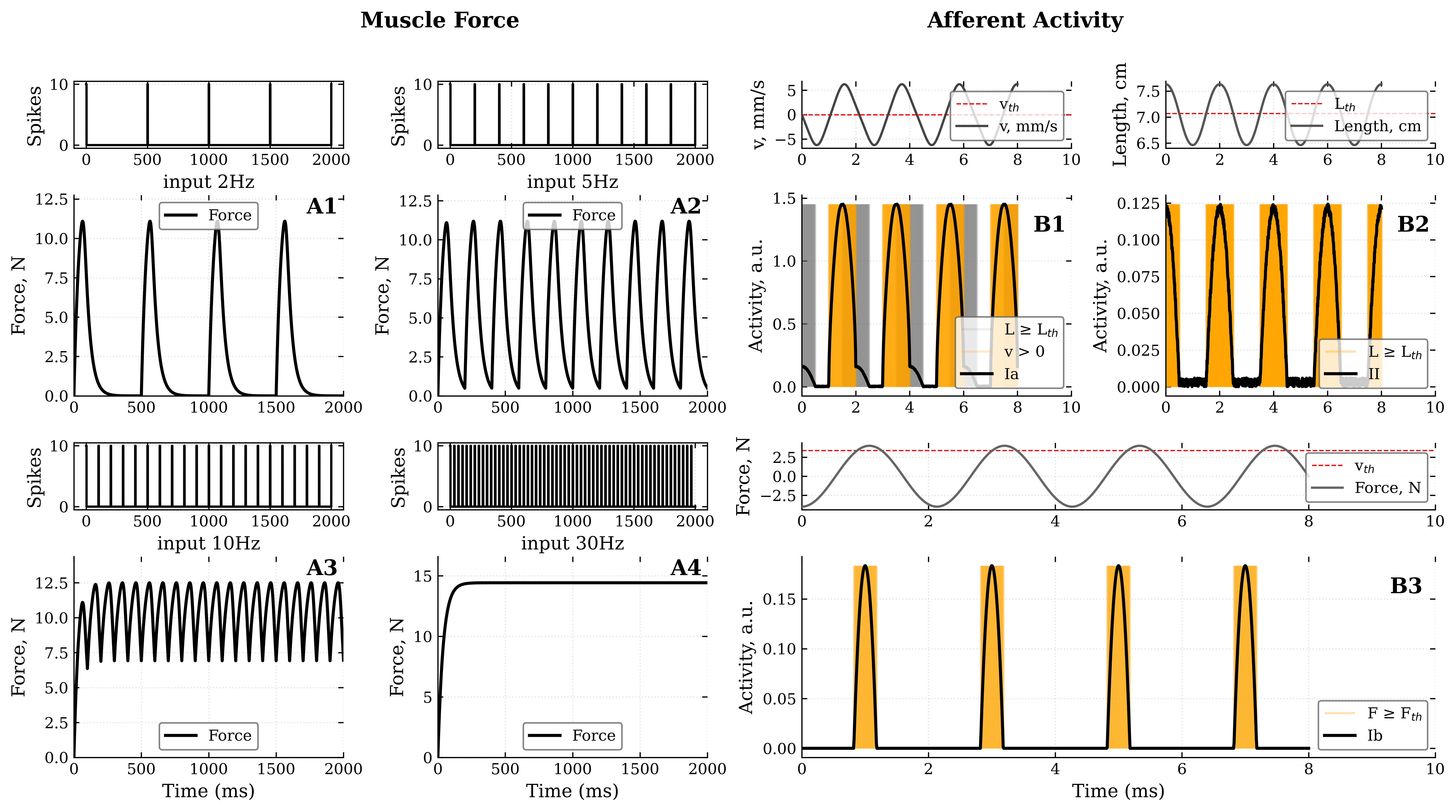}
    \caption{Muscle force and afferent activity. (\textbf{A1}--\textbf{A4}) Muscle force response to stimulation at different frequencies (2 Hz, 5 Hz, 10 Hz, 30 Hz). The top row displays input spike trains, and the bottom row shows the resulting muscle force. (\textbf{B1}--\textbf{B3}) Activity of muscle afferents. (\textbf{B1}) Ia afferent activity depending on velocity ($v$) and length ($L$). (\textbf{B2}) II afferent activity depending on length ($L$). (\textbf{B3}) Ib afferent activity depending on force ($F$). Red dashed lines indicate thresholds ($v_{th}$, $L_{th}$, $F_{th}$).}
    \label{fig:comprehensive_muscle_afferent_gridspec}
\end{figure*}

Following Wilson et al. \cite{Wilson2011}, muscle force $F$ is derived from motoneuron activity $f(v)$ using a simplified model combining linear ODEs and a static nonlinearity. The governing equations are:

\begin{equation}
    \begin{cases}
        \dfrac{dC_n}{dt} = W \cdot f(v) - \dfrac{C_n}{\tau_c},\\[8pt]
        x = \dfrac{C_n^m}{C_n^m + k^m},\\[8pt]
        \dfrac{dF}{dt} = A x - \dfrac{F}{\tau_1},
    \end{cases}
    \label{eq:Wilson}
\end{equation}

where, $C_n$ is the normalized synaptic activation, $W=0.5$ the synaptic weight, and $\tau_c = 21~\text{ms}$, $\tau_1 = 39~\text{ms}$ the synaptic and force relaxation constants. The nonlinear function $x \in [0,1]$ ($m=3$, $k=0.5$) captures excitation--contraction saturation, while $A = 0.37~\text{N/ms}$ scales neural drive to force. This structure accurately reproduces phase-locked muscle responses across contraction types and frequencies (Fig. \ref{fig:comprehensive_muscle_afferent_gridspec}(\textbf{A1-4})).

\label{subsec:afferents}

To close the sensorimotor loop, the model incorporates primary, secondary, and force-dependent afferent pathways for both flexor (Ia-F, Ib-F, II-F) and extensor (Ia-E, Ib-E, II-E) muscles(Fig. \ref{fig:cpg}). Afferent firing rates are computed using a modified version of the empirical relations proposed by Prochazka et al. \cite{Prochazka1979} adapted by S. Markin \cite{Markin2010}:

\begin{equation}
    \begin{cases}
        Ia = k_v \cdot v_{\text{norm}}^{p_v} + k_{dI} \cdot d_{\text{norm}} + k_{nI} \cdot f(v),\\[6pt]
        Ib = k_F \cdot F_{\text{norm}},\\[6pt]
        II = k_{dII} \cdot d_{\text{norm}} + k_{nII} \cdot f(v),
    \end{cases}
    \label{eq:Prochazka}
\end{equation}

\noindent where $v_{\text{norm}} = v / L_{th}$ is the normalized muscle contraction velocity, $p_v = 0.6$ accounts for nonlinear velocity sensitivity, and $d_{\text{norm}} = \max\bigl(0, (L - L_{th})/L_{th}\bigr)$ represents normalized muscle length relative to the activation threshold $L_{th} = 59~\text{mm}$. The normalized force is defined as $F_{\text{norm}} = \max\bigl(0, (F - F_{th})/F_{\text{max}}\bigr)$ with force threshold $F_{th} = 3.38~\text{N}$. The scaling coefficients are set as follows: $k_v = 6.2$, $k_{dI} = 2.0$, $k_{dII} = 1.5$, $k_{nI} = k_{nII} = 0.06$, and $k_F = 1.0$. These relations ensure physiologically plausible afferent signaling that dynamically modulates CPG phase duration and amplitude in response to load and kinematic changes.

Neuronal activity within the CPG network is simulated using the Izhikevich spiking neuron model \cite{Izhikevich2003}, which balances biological realism with computational efficiency. The membrane potential $v_i$ and recovery variable $u_i$ of the $i$-th neuron evolve according to:

\begin{equation}
    \begin{cases}
        \dfrac{dv_i}{dt} = 0.04 v_i^2 + 5 v_i + 140 - u_i + I^{\text{app}}_i + \sum_{j} I^{\text{syn}}_{i,j} + I^{\text{aff}}_i,\\[8pt]
        \dfrac{du_i}{dt} = a_i (b_i v_i - u_i),
    \end{cases}
    \label{eq:izhikevich}
\end{equation}

\noindent with the discontinuous reset condition applied when $v_i \geqslant v_{\text{peak}}$ ($30~\text{mV}$):

\begin{equation}
    v_i \leftarrow c_i, \quad u_i \leftarrow u_i + d_i.
    \label{eq:izhikevich_reset}
\end{equation}

Here, $I^{\text{app}}_i$ represents external drive current, $I^{\text{syn}}_{i,j}$ denotes the synaptic input from presynaptic neuron $j$, and $I^{\text{aff}}_i$ aggregates afferent feedback currents. The parameters $a_i$, $b_i$, $c_i$, and $d_i$ are tuned to reproduce specific electrophysiological phenotypes. In this study, neurons are configured as follows: regular-spiking (RS) for MN and R-In; fast-spiking (FS) for Ia-In and Ib-In; and chattering/bursting (CH) for CPG-N and CPG-In, following the parameter sets established in \cite{Izhikevich2003}. For CPG-N neurons, reducing the Izhikevich recovery parameter to $a=0.005$ slows adaptation dynamics, prolonging interburst intervals and lowering burst frequency to stabilize the half-center rhythm.

Synaptic transmission is modeled as an exponentially decaying current triggered by presynaptic spikes:

\begin{equation}
    \dfrac{dI^{\text{syn}}_{i,j}}{dt} = w_{i,j} \cdot \delta(t - t^{\text{spike}}_j) - \dfrac{I^{\text{syn}}_{i,j}}{\tau_{i,j}},
    \label{eq:synaptic_current}
\end{equation}

\noindent where $w_{i,j}$ is the synaptic weight (positive for excitation, negative for inhibition), $t^{\text{spike}}_j$ is the spike time of the presynaptic neuron, and $\tau_{i,j}$ is the synaptic decay time constant.

Afferent feedback currents are linearly summed into each neuron:

\begin{equation}
    I^{\text{aff}}_i = \sum_{k} p_{i,k} \, fb_k,
    \label{eq:afferent_current}
\end{equation}

\noindent where $p_{i,k}$ denotes the synaptic weight of the $k$-th afferent pathway $fb_k$ projecting onto neuron $i$. This closed-loop circuit generates its own rhythm, which can be modulated by sensory input. This allows for the study of antiphase synchronization --- which will be discussed in the following sections.

\subsection{Experimental Modeling Protocols}
\label{sec:protocols}

To systematically investigate the adaptive dynamics of the locomotor CPG network, we designed two complementary experimental paradigms. We use a single computational framework for both protocols. In the first, we vary the external drive while fixing synaptic weights; in the second, we do the opposite---varying weights while keeping the drive fixed.

All simulations were performed with a total duration of $10$~s and a sampling rate of $5$~points/ms ($5$~kHz). To exclude transient dynamics, the first $5$~s of each simulation were discarded from subsequent analysis.

To simulate natural "neural noise" --- random activity from neighboring cells --- we added weak Gaussian noise $\xi(t)$ to the input signals. The noise magnitude was strictly bounded to $5\%$ of the primary input amplitude $A$:

$|\xi(t)| \leq 0.05 \cdot A.$
Unless otherwise specified in a given protocol, all recurrent synaptic weights within the CPG microcircuit were fixed at $w = 0.5$ to establish a symmetric baseline operating point. The default topology and weights of sensory afferent pathways to flexor and extensor half-centers are summarized in Fig.~\ref{fig:cpg}.

The first protocol is based on changing the parameters of the input current to the neural system at fixed synaptic weights. In this protocol, the internal CPG architecture and afferent weights were held constant ($w_{\text{internal}}=0.5$). External drive was applied exclusively to the rhythm-generating neurons (CPG-N and CPG-IN) and distributed identically within each half-center. Three input modalities were investigated:

The tonic drive follows:
\begin{equation}
I_{\text{DC}}(t) = I_0 + \xi(t),
\end{equation}
where $I_0 \in [0, 20]$~nA with a step of $0.5$~nA. By trying all possible pairs of input currents to the flexor and extensor sides, we can fully explore how changes in drive affect the network's output.

Both oscillatory modalities share common control parameters: amplitude, period $T$, and inter-half-center phase shift $\varphi$. Amplitudes were fixed at $A_{\text{CPG-IN}} = 5$~nA and $A_{\text{CPG-N}} = 4$~nA:

\begin{align}
I_{\text{CPG-N}}(t) &= 4 \cdot f(t) + \xi(t), \\
I_{\text{CPG-IN}}(t) &= 5 \cdot f(t) + \xi(t),
\end{align}
where $f(t)$ denotes the base waveform. For sinusoidal inputs:
\begin{equation}
I_{\text{sin}}(t) = A \cdot \sin\left(\frac{2\pi t}{T} + \varphi\right) + \xi(t).
\end{equation}
For rectangular pulses:
\begin{equation}
I_{\text{rect}}(t) =
\begin{cases}
A + \xi(t), & \text{if } (t \bmod T) < \tau, \\
\xi(t), & \text{otherwise},
\end{cases}
\end{equation}
where $\tau$ is the pulse duration.

Parameter sweeps were constrained to physiologically and computationally valid regimes ($\tau \leq T$). Periods $T$ were distributed equally between $1$~ms and $1$~s (5 values for meander, 8 for sinusoid). Phase shifts $\varphi$ spanned $[0, 2\pi]$ (7 equidistant points for meander, 8 for sinusoid) to capture in-phase and anti-phase coordination. For meander stimuli, $\tau$ was additionally sampled at 3 equidistant points within $[0, 400]$~ms.

The second protocol is based on the variation of synaptic weights between afferent neurons and half-center neurons that set the rhythm. This protocol isolates the impact of network connectivity on rhythmogenesis by fixing the input drive (typically at baseline DC or a predefined oscillatory pattern) and systematically varying synaptic strengths. Synaptic weights within the four reciprocal inhibitory circuits (CPG-N, Ib-IN, Ia-IN, and R-IN) were varied across the discrete set $\{0.001, 0.5, 1.0\}$, representing near-disconnection, weak coupling, and strong inhibition, respectively. All possible combinations of these weight values were tested across the target circuits to explore the full parameter space of inhibitory connectivity. Additionally, the influence of muscle feedback was probed by independently scaling pairs of homologous afferent pathways (Ia, II, Ib) for flexor and extensor sides across the same weight set $\{0.001, 0.5, 1.0\}$, while maintaining default afferent topology per Fig.~\ref{fig:cpg}. This design allows us to isolate the contribution of length-dependent (Ia, II) and force-dependent (Ib) feedback to phase transitions and speed control, independent of the intrinsic CPG drive.

To quantify the stability of antiphase synchronization between antagonistic neuronal groups, we compute the Phase-Locking Value (PLV) using the Hilbert transform \cite{Lachaux1999, Cohen1995}.

In the time domain, applying the Hilbert transform $\mathcal{H}\{s(t)\}$ to a real-valued signal $s(t)$ is equivalent to convolving the original signal with the function $\frac{1}{\pi t}$ \cite{Oppenheim2010}:
\begin{equation}
    \mathcal{H}\{s(t)\} = s(t) * \frac{1}{\pi t} = \frac{1}{\pi} \text{P.V.} \int_{-\infty}^{\infty} \frac{s(\tau)}{t - \tau} \, d\tau,
    \label{eq:hilbert_time}
\end{equation}
where P.V. denotes the Cauchy principal value. Equivalently, in the frequency domain, the Hilbert transform corresponds to multiplying the signal's Fourier transform $S(f)$ by $-i \cdot \text{sgn}(f)$, where $\text{sgn}(f)$ is the sign function \cite{Oppenheim2010, Cohen1995}:
\begin{equation}
    \mathcal{F}\{\mathcal{H}\{s(t)\}\} = -i \cdot \text{sgn}(f) \cdot S(f).
    \label{eq:hilbert_freq}
\end{equation}

The combination of the original real signal and its Hilbert transform (as the imaginary part) creates a complex analytic signal $z(t)$ \cite{Picinbono1997, Cohen1995}. For two bandpass-filtered signals $s_1(t)$ and $s_2(t)$ (1--10~Hz), the analytic representations are constructed as:
\begin{equation}
    z_k(t) = s_k(t) + i\,\mathcal{H}\{s_k(t)\}, \qquad k \in \{1,2\}.
    \label{eq:analytic_signal}
\end{equation}
This analytic representation enables the precise extraction of the instantaneous amplitude (envelope) $A_k(t) = |z_k(t)|$ and the instantaneous phase $\phi_k(t)$ of the signal \cite{Cohen1995, Picinbono1997}. The instantaneous phases $\phi_k(t)$ are explicitly extracted as the argument of the complex analytic signal:
\begin{equation}
    \phi_k(t) = \arg\!\bigl[z_k(t)\bigr]
    = \operatorname{arctan}\!\bigl(\operatorname{Im}[z_k(t)],\, \operatorname{Re}[z_k(t)]\bigr),
    \label{eq:phase_extraction}
\end{equation}
where $\operatorname{Re}[\cdot]$ and $\operatorname{Im}[\cdot]$ denote the real and imaginary parts, respectively, and $\operatorname{arctan2}(\cdot,\cdot)$ ensures quadrant-correct phase values in $(-\pi,\pi]$. The instantaneous phase difference is then defined as
\begin{equation}
    \Delta\phi(t) = \phi_1(t) - \phi_2(t),
    \label{eq:phase_diff}
\end{equation}
and the Phase-Locking Value (PLV) is computed as the magnitude of the mean resultant vector:
\begin{equation}
    \mathrm{PLV} = \left| \frac{1}{N} \sum_{n=1}^{N} \exp\!\bigl[i\,\Delta\phi(t_n)\bigr] \right|,
    \label{eq:plv_hilbert}
\end{equation}
where $N$ is the number of samples in the analysis window and $i$ is the imaginary unit.
\section{Results}\label{sec2}

\subsection{Classification of Dynamic Regimes and Validation of the Composite Metric}\label{subsec:regimes_metric}

Numerical simulations of the biomorphic spiking CPG network revealed four distinct dynamic regimes characterizing the output muscle activity.

\begin{figure*}[t]
\centering
    \includegraphics[width=\textwidth]{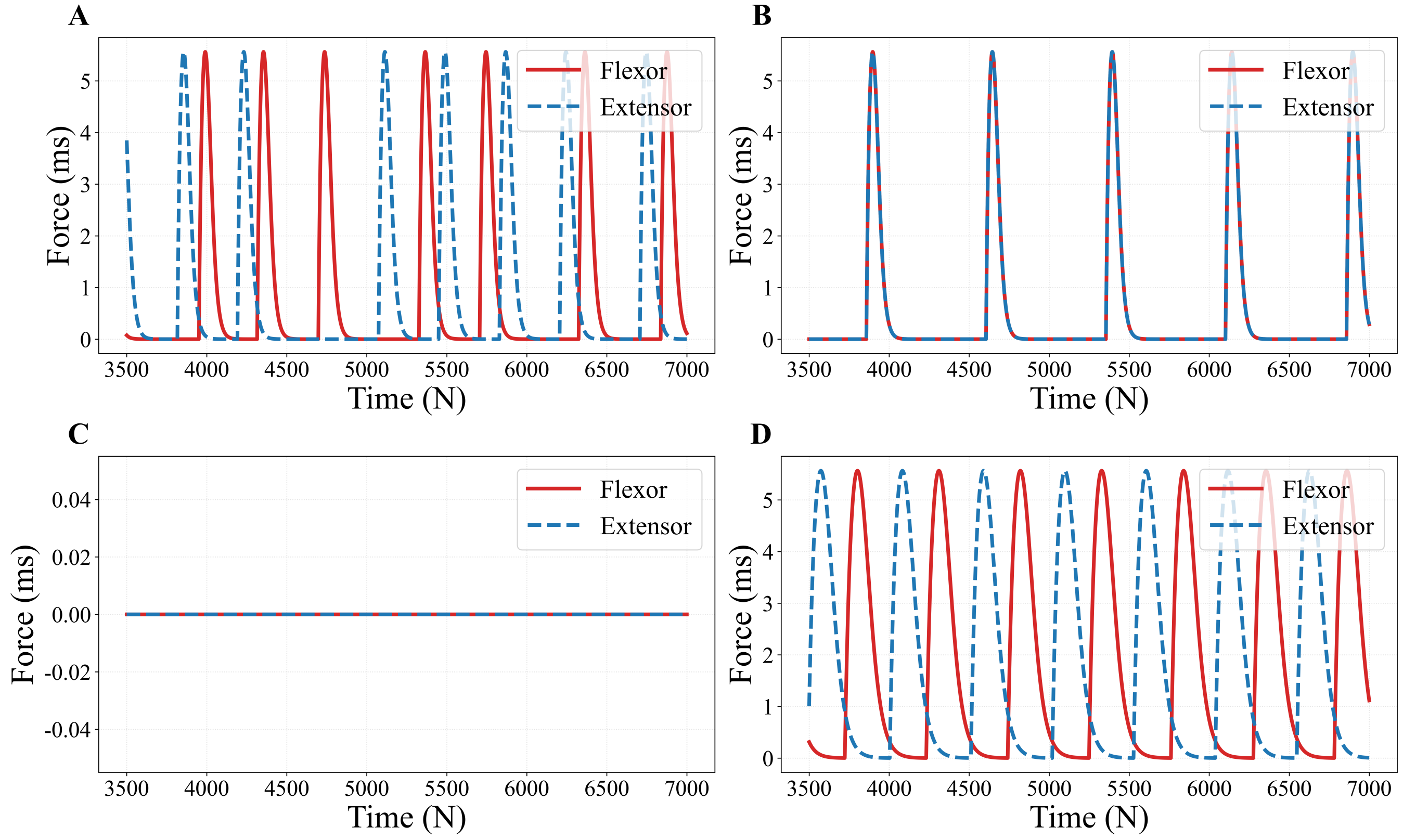}
\caption{Examples of the four primary dynamic regimes of CPG output muscle activity. Top row: temporal force realizations for flexors ($F_{\text{flex}}$, blue) and extensors ($F_{\text{ext}}$, red). Bottom row: corresponding phase portraits obtained via Hilbert transform in the 1--10~Hz band.}
\label{fig:regimes_examples}
\end{figure*}

The phase desynchronization regime is characterized by the presence of rhythmic activity that lacks stable phase relationships between antagonistic groups, potentially modeling pathological or transitional states such as ataxia, gait instability, or adaptive reconfiguration of locomotor circuits during speed changes \cite{bib77} Fig.~\ref{fig:regimes_examples}A.
In the full in-phase synchronization regime, flexor and extensor groups are activated simultaneously, which models pathological co-contraction states such as spasticity or rigidity in upper motor neuron syndromes, where antagonistic muscles activate concurrently to stiffen joints and stabilize posture under neurological impairment \cite{bib67} Fig.~\ref{fig:regimes_examples}B.
The no-oscillation regime manifests as a suppression of rhythmogenesis, resulting in the absence of periodic muscle activity and producing a stationary or noise-like signal; this regime may correspond to conditions such as spinal shock, pharmacological blockade of synaptic transmission, or complete motor inhibition \cite{bib16} Fig.~\ref{fig:regimes_examples}C.
Finally, the anti-phase synchronization regime features a stable alternation of flexor and extensor group activities, corresponding to normal locomotor patterns such as walking or swimming, where antagonistic muscle groups activate in strict alternation to produce efficient rhythmic movement \cite{bib16} Fig.~\ref{fig:regimes_examples}D.
Representative temporal realizations of muscle forces and corresponding phase portraits for each regime are shown in Fig.~\ref{fig:regimes_examples}.

To automatically classify these regimes according to our requirements, we develop a heuristic composite metric $S_{\text{composite}} \in [0,1]$ that captures the distinctive features of each dynamic state:

\begin{equation}
S_{\text{composite}} = \frac{w_{\text{plv}} \cdot \text{PLV} + w_{\text{amp}} \cdot A + w_{\text{ph}} \cdot P_h}{w_{\text{plv}} + w_{\text{amp}} + w_{\text{ph}}} \cdot g(P),
\label{eq:composite_metric}
\end{equation}

where the components represent: (1)~phase-locking value via Hilbert transform in the 1--10~Hz band, (2)~amplitude balance between flexor and extensor neuronal groups, and (3)~phase shift proximity to $180^\circ$. The gating function $g(P) = \exp(-\alpha \max(0, P-2))$ applies a soft penalty when the envelope peak count $P = \max(P_{\text{flex}}^{\text{env}}, P_{\text{ext}}^{\text{env}})$ exceeds 2, with $\alpha$ optimized jointly with the weights.

An expert-annotated dataset comprising $N = 7000$ simulation segments is compiled to validate the metric. On this dataset, we train and evaluate both the baseline Hilbert PLV metric and our proposed composite metric in a binary classification task (\texttt{anti\_phase} vs \texttt{not\_anti\_phase}).

The optimized composite metric achieves an AUC of $0.766 \pm 0.009$, demonstrating a statistically significant improvement over the baseline PLV approach ($\text{AUC} = 0.741 \pm 0.011$). The optimization procedure reveals that amplitude balance dominates the discriminative power ($w_{\text{amp}} \approx 0.963$), which aligns with biomechanical requirements for symmetric locomotor patterns. Consequently, the proposed metric enables robust, interpretable detection of anti-phase regimes, fulfilling the computational efficiency needed for online adaptation in robotic control systems.

\subsection{Influence of Stimulation Parameters and Synaptic Weights on Dynamic Regimes}\label{sec:param_influence}

The stability of antiphase synchronization is governed by the joint influence of external stimulation, synaptic architecture, afferent feedback, and neural noise.

To disentangle their contributions, we applied variance analysis and machine learning techniques.

Two-way ANOVA reveals statistically significant main effects of stimulation type and mean synaptic weight, as well as a strong interaction between them.

Stimulation type exerts the dominant effect, reflecting the hierarchical organization of locomotor control: descending pathways set pacing, while spinal microcircuits coordinate phase alternation.

Under tonic drive (DC), the network operates as an autonomous generator where reciprocal inhibition stabilizes anti-phase locking.

Periodic stimulation introduces a conflict with the endogenous rhythm, producing nonlinear saturation and context-dependent efficacy of synaptic tuning, consistent with biological phase-resetting mechanisms.

A clear hierarchy of stimulation efficacy emerges: DC yields the highest and most stable synchronization, whereas periodic stimuli (Square, Sine) produce lower median values and bimodal distributions.

Under DC, strengthening mutual inhibition monotonically improves synchronization, aligning with classical half-center oscillator theory.

\begin{figure*}[t]
\centering
\includegraphics[width=\textwidth]{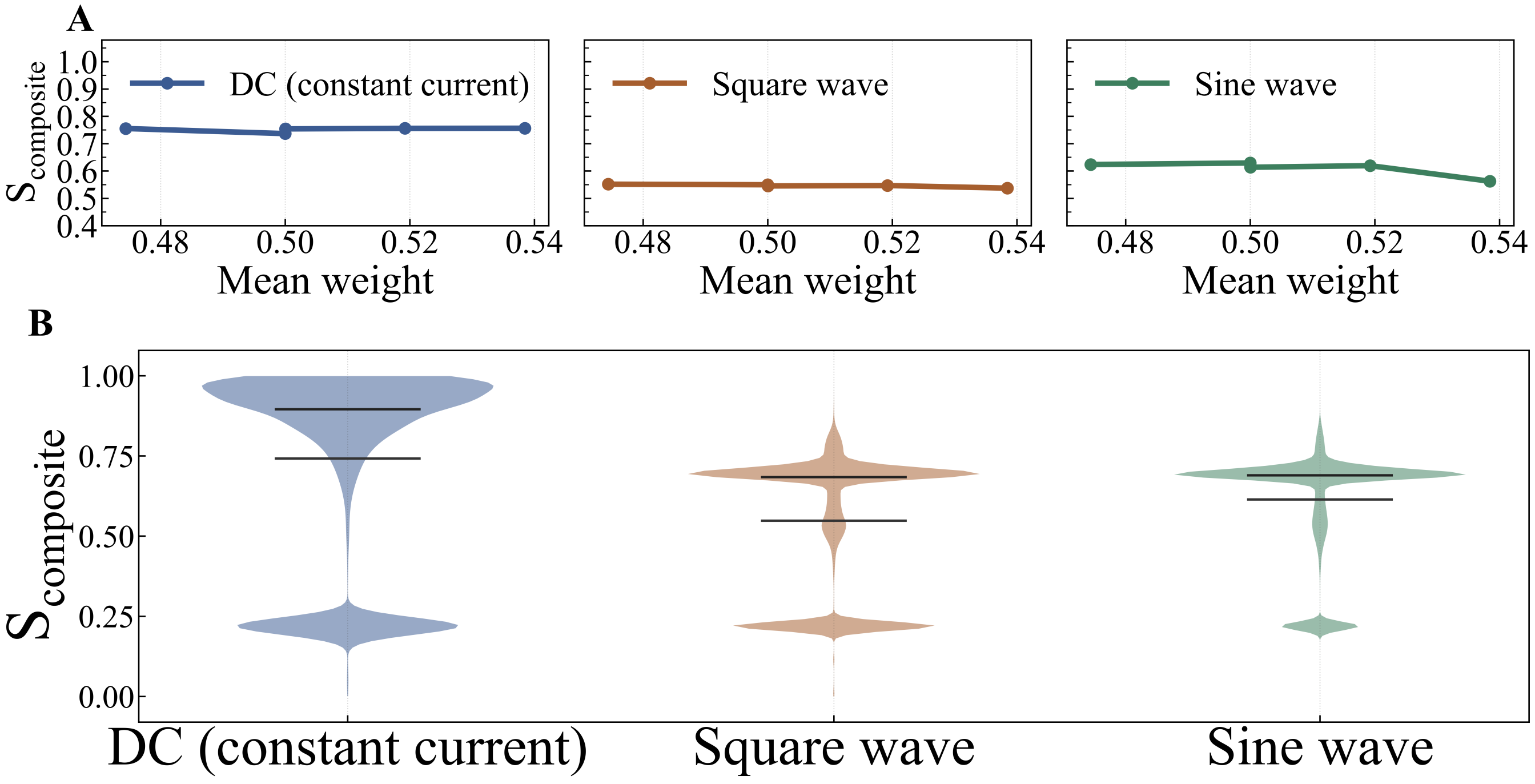}
\caption{\textbf{Statistical validation of anti-phase synchronization dependence on stimulation type and synaptic weights in the biomorphic spiking network.}
\textbf{Panel (a)} shows the interaction between stimulation type and mean synaptic weight for the composite phase-locking metric $S_{\text{composite}}$: dynamic inputs (Square, Sine) exhibit nonlinear saturation with plateau at intermediate weights, indicating context-dependent efficacy of synaptic tuning ($F_{2,1496928}=236.5$, $p<0.001$ for interaction).
\textbf{Panel (b)} displays the distribution of $S_{\text{composite}}$ values across stimulation types (violin plot): DC achieves higher median synchronization (0.895, IQR: 0.71--0.97) with narrower spread versus periodic stimuli (Sine: median = 0.690, IQR: 0.60--0.77; Square: median = 0.684, IQR: 0.58--0.76), confirming systematic superiority of tonic drive; horizontal lines within violins indicate median and quartiles.}
\label{fig:anova_three_panel}
\end{figure*}

These dynamics are illustrated in Fig.~\ref{fig:anova_three_panel}.

Notably, in addition to the primary synchronization clusters, Panel~(b) of Fig.~\ref{fig:anova_three_panel} reveals a distinct secondary peak centered at $S_{\text{composite}} \approx 0.250$. This low-value mode corresponds to the \textit{no-oscillation} regime, characterized by the absence of periodic muscle activity and near-zero phase-locking ($\text{PLV} < 0.100$). Crucially, this regime manifests consistently across all stimulation modalities (DC, Sine, Square), forming a well-separated cluster that does not overlap with rhythmic states. The composite metric thus provides explicit discrimination of non-rhythmic dynamics without requiring manual threshold selection, enabling automated regime classification in closed-loop neuromechanical applications.

Permutation importance analysis via Random Forest identifies a consistent synaptic hierarchy: Ia interneurons (reciprocal inhibition) show the highest normalized importance, followed by CPG core connections, Renshaw pathways, and phase-switching projections.

A strong negative correlation between the CPG\_N\_Flex$\to$CPG\_N\_Ext weight and PLV confirms the fundamental role of reciprocal inhibition in generating anti-phase patterns.

It is important to note that the importance of synaptic pathways depends on context: rectangular stimulation and sinusoidal stimulation were heavily dependent on the CPG core, while direct current distributed importance between inhibition and feedback circuits.

Although synaptic weights consistently contribute detectable effects across all configurations, stimulation type remains the primary regime determinant.

These findings support a hierarchical control framework in which hierarchy is achieves through the dominant influence of descending signals that define the operational context, whereas the synaptic architecture---specifically, the synaptic weights---of spinal circuits ensures robust pattern implementation via dynamic modulation~\cite{bib53}.

\begin{figure*}[t]
\centering
\includegraphics[width=\textwidth]{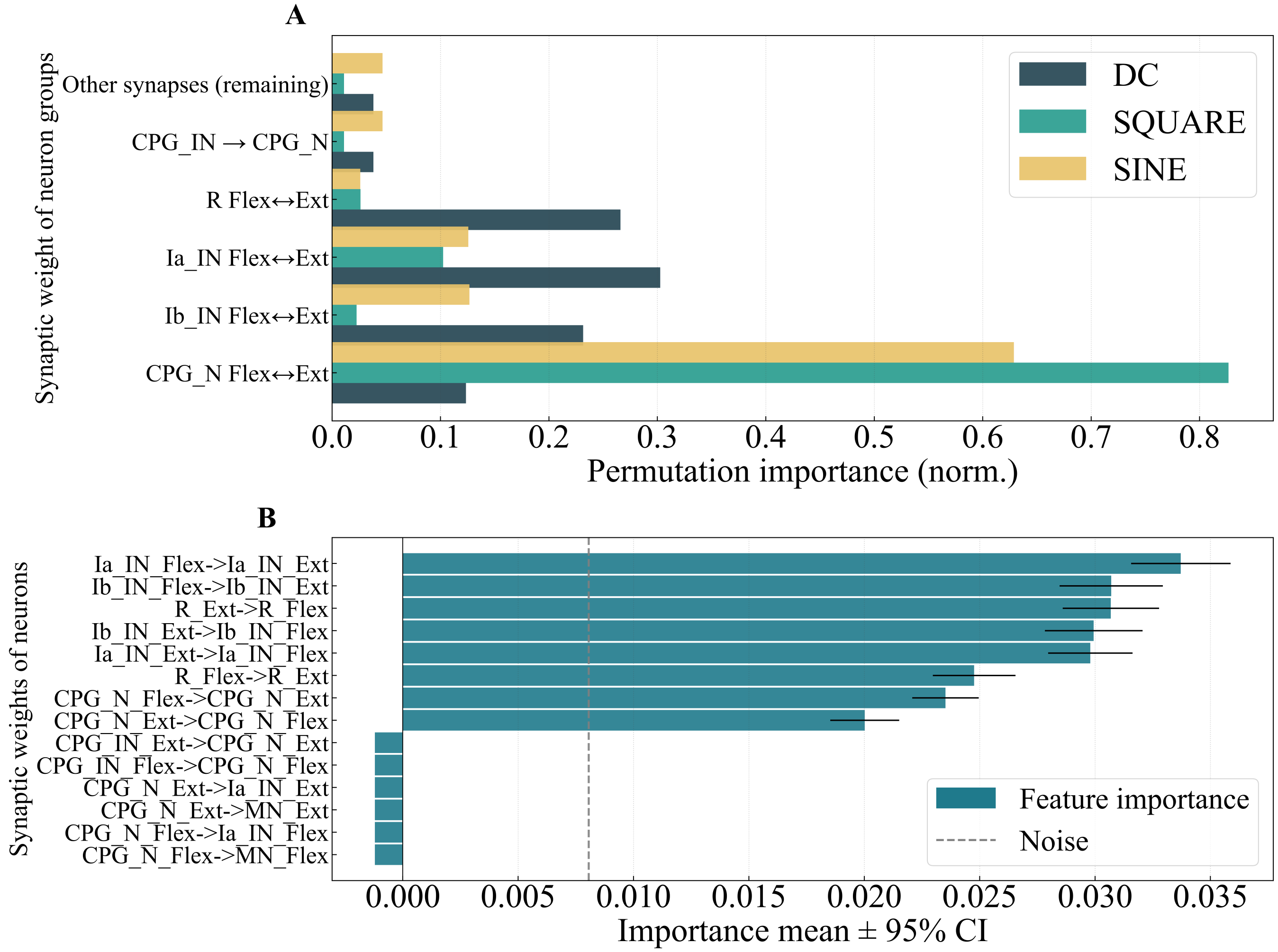}
\caption{\textbf{Analysis of synaptic determinants of anti-phase synchronization.}
(\textbf{a})~Context-dependent importance profiles reveal that the relative contribution of functional connectivity groups shifts with stimulation waveform, supporting a hierarchical, modular organization.
(\textbf{b})~Subsets of synaptic weight invariably demonstrate a statistically significant effect on synchronization, which allows us to analyze not only the influence of neuronal groups, but also individual connections.
The error columns in (\textbf{b}) represent 95\% confidence intervals calculated using permuted resampling ($n_{\text{repeats}} = $30 per repetition, summed over $k=3$ stratified CV multiplicities). Features with confidence intervals exceeding zero are classified as statistically indistinguishable from noise.}
\label{fig:rf_megapanel}
\end{figure*}

These trends are summarized in Fig.~\ref{fig:rf_megapanel}.

\section{Discussion}\label{sec12}

The present study demonstrates that stable antiphase synchronization in biomorphic spiking central pattern generators (CPGs) emerges from a hierarchical interplay between external drive characteristics and intrinsic synaptic architecture. Central to this investigation is the development of a closed-loop CPG architecture with explicit muscle output pathways and biomorphic proprioceptive feedback. By embedding the network within a neuromechanical loop, we bypass the need to monitor full neural state variables and instead classify dynamic regimes directly from muscle force dynamics, providing a functionally grounded assessment of locomotor stability. The dominant effect of stimulation waveform aligns with neurophysiological evidence that descending pathways primarily set locomotor frequency and amplitude, while spinal microcircuits handle phase coordination and reflex modulation. Under tonic (DC) drive, reciprocal inhibitory strength contributes to rhythm robustness and phase alternation, consistent with classical half-center oscillator models \cite{Brown1911,Matsuoka1985}. Conversely, periodic inputs induce nonlinear saturation, reflecting forced oscillator dynamics and phase-locking constraints typical of coupled neural systems subject to external pacing.

The proposed composite synchronization metric ($S_{\text{composite}}$) addresses a critical gap in locomotor modeling: traditional phase-only measures fail to capture amplitude symmetry and temporal stability, both essential for biomechanical gait transitions and robotic control \cite{Ijspeert2008}. By integrating phase coherence, flexor-extensor force balance, and soft peak-count penalties, the optimized metric achieves robust anti-phase detection (AUC = 0.766) while preserving biophysical interpretability, offering a transparent alternative to black-box classifiers that require extensive labeled datasets \cite{Roberts2017}.

Despite these advances, several limitations warrant consideration. The model implements a fixed descending drive from higher nervous centers to the CPG, treating this input as a constant tonic or periodic signal without accounting for activity-dependent plasticity in these pathways. In biological systems, corticospinal and reticulospinal projections exhibit spike-timing-dependent plasticity (STDP) and homeostatic adaptation, which are crucial for motor learning, developmental refinement, and recovery after injury. By holding descending weights constant, our framework captures steady-state locomotor patterns but does not address how the network adapts to novel terrains, learns new gaits, or compensates for perturbations through synaptic reorganization. The phenomenological muscle model captures essential force dynamics but omits detailed biomechanical coupling and tendon compliance, which can influence reflex latency and phase resetting \cite{Wilson2011}. Additionally, additive Gaussian noise approximates background neural activity but does not model ion-channel stochasticity or hardware-level variability inherent in neuromorphic deployments \cite{Davies2021}. Future model development will integrate plasticity rules in descending pathways, embed the CPG within reinforcement learning frameworks for adaptive gait modulation \cite{Eshraghian2022}, and validate the architecture on event-driven neuromorphic processors to assess energy-efficient real-time control \cite{Pehle2022}.

\section{Conclusion}\label{sec13}

This work establishes a statistically validated framework for identifying and stabilizing antiphase synchronization regimes in biomorphic spiking neural networks. Through comprehensive numerical experimentation and multivariate analysis, we demonstrate that stable alternating bursting emerges not from synaptic weights alone, but from their context-dependent interaction with stimulation waveform. Tonic drive yields the most robust phase-locking, while periodic inputs require precise weight tuning to avoid dynamical conflict between endogenous rhythmogenesis and external forcing. The optimized composite metric bridges the gap between abstract phase synchronization measures and biomechanically grounded locomotor control, offering a transparent, statistically rigorous alternative to purely data-driven classifiers. These findings provide actionable parametric guidelines for designing closed-loop neuromechanical controllers and inform the development of adaptive neuroprosthetic interfaces. Future integration with neuromorphic hardware and plasticity-driven learning will enable scalable, energy-efficient robotic locomotion systems that dynamically adapt to environmental perturbations.

\section*{Acknowledgements}

This work was supported by the Ministry of Science and Higher Education of the Russian Federation. The modeling of the spiking neural network was carried out under the federal assignment (project No. FSMG-2025-0070), while the data analysis and interpretation of the results were supported by project No. FSMG-2024-0047.

 \section*{Declarations}

The authors declare that they have no known competing financial interests or personal relationships that could have appeared to influence the work reported in this paper.

\section*{DATA AVAILABILITY}
The data that support the findings of this study are available from the corresponding author upon reasonable request.

\section*{AUTHOR CONTRIBUTIONS}
Artem Guba: Conceptualization, Formal analysis, Writing - original draft.
Nikolai Kovalev: Methodology, Software, Validation.
Alexey Zharinov: Investigation, Data curation.
Vyacheslav Demin: Resources, Supervision.
Viktor B. Kazantsev: Conceptualization, Writing - review \& editing.

\bibliography{references}

\end{document}